\documentstyle[preprint,aps,epsf,floats]{revtex}
\tighten

\newcommand{\lsim}{\raisebox{-0.7ex}{$\stackrel{\textstyle <}{\sim}$ }}

\begin{document}

\preprint{DOE/ER/40427-32-N96}
\title{SU(3) Breaking in Neutral Current Axial Matrix Elements \\
and the Spin-Content of the Nucleon}
\author{Martin  J. Savage
\footnote{{\tt savage@thepub.phys.washington.edu}}
and James Walden
\footnote{{\tt walden@phys.washington.edu}}
 }
\address{
Department of Physics, University of Washington,
\\
Seattle, WA
98195
}
\maketitle

\begin{abstract}

We examine the effects of SU(3) breaking in the  matrix
elements of the flavour-diagonal axial currents
between octet baryon states.
Our calculations of $K, \eta$ and $\pi$ loops
indicate that the SU(3) breaking may be
substantial for some matrix elements and at the very
least indicate large uncertainties.
In particular, the strange axial matrix element in the
proton determined from the measurements of  $g_1(x)$
is found to have  large uncertainties and might yet be zero.
We estimate the strange axial matrix element
in the proton to be
$-0.35\  \lsim\  \Delta s\  \lsim\ 0$ and the matrix element of
the flavour-singlet current in the proton
to be
$-0.1\  \lsim\  \Sigma\  \lsim\ +0.3$
from the E-143 measurement of
$\int dx\ g_1(x)\ =\ 0.127\pm 0.004 \pm 0.010$~.

The up-quark content of the $\Xi^-$ is discussed and
its implications for nonleptonic weak processes
discussed.
We also estimate the matrix element of the axial current
coupling to the $Z^0$ between all octet baryon states.
This may be important for neutrino interactions in dense
nuclear environments, where hyperons may play an important role.

\end{abstract}

\bigskip
\vskip 4.0cm
\leftline{October 1996}

\vfill\eject

One of the more exciting realizations in hadronic
physics of the last few
years is that the strange quark may play
an important role in the structure of the nucleon
\cite{KM} .
While this may seem somewhat unnatural in the
context of the most naive
quark model, it is
perfectly natural from the standpoint of QCD.
Matrix elements of the strange-vector
current must vanish at
zero-momentum transfer
between states with zero net-strangeness, however, matrix
elements of the axial
current need not.
Recent measurements suggest that the matrix element of the
strange axial current in the proton is
$\Delta s = -0.12\pm 0.04$
\cite{SMCd}~.
In addition, one would like to know what fraction of the
nucleon spin is carried by the quarks themselves, which is equivalent
to determining the matrix
element of  flavour singlet axial current in the proton, $\Sigma$~.
This is, of course, intimately related to the matrix element
of the strange axial current and
present analysis suggests that
$\Sigma = 0.2\pm 0.1$  \cite{SMCd}~,
much smaller than the quark model estimate of $\Sigma\sim 0.58$.
There have been intense  theoretical and
experimental efforts to
extract $\Delta s$ and $\Sigma$ to address the present ``spin-crisis''
and such efforts
continue (for recent reviews, see \cite{Jafa,MDDPKB})~.

A vital ingredient in the present
determination of  $\Sigma$ and $\Delta s$
is the
matrix element of the
$j_5^{\mu, 8} = \overline{u}\gamma^\mu\gamma_5 u
+ \overline{d}\gamma^\mu\gamma_5 d
- 2 \overline{s}\gamma^\mu\gamma_5 s $
axial current in the nucleon, which
cannot be measured directly
but must be inferred
from the approximate SU(3) symmetry observed in nature.
The question of SU(3) breaking in the matrix element
of the $j_5^{\mu, 8}$ current
has been previously addressed
\cite{LL,ES,DDJM}.
In \cite{ES}\  it was assumed that the breaking
in the matrix elements of the axial currents
was proportional to the
breaking in the octet baryon masses and in \cite{LL}\
a model of the SU(3)
breaking was employed.
A more systematic approach was that of \cite{DDJM}\ in which
the breaking was analyzed in the
context of the large-$N_C$ limit of QCD.
It was found that the matrix element of the $j_5^{\mu, 8}$
axial current was
substantially reduced from its value in the symmetry limit.

In the limit of flavour SU(3) symmetry the three light quark
contributions to the nucleon
axial matrix elements are uniquely
determined by three low energy observables.
In this limit, two of these observables, $F$ and $D$,
can be extracted from
nuclear $\beta$-decay and from
the semileptonic decay of strange hyperons.
The third experimental constraint comes from a measurement of
the axial singlet current in
the nucleon, presently accomplished by measuring the $g_1(x)$ spin
dependent structure function of the nucleon
\cite{SMCg,SLACg}\
and using the SU(3) symmetry to remove the flavour
octet contributions.
In the real world we know that this symmetry is only approximate,
broken by the
difference between the mass of
the strange quark and of the up and down quarks.
Each of the matrix elements of the octet and singlet axial
currents will receive SU(3) breaking contributions, with the
leading contributions having the
form $m_s \log m_s$ followed by terms of the form $m_s$ and higher.
The leading contributions with non-analytic dependence on $m_s$
arise from hadronic kaon loops while
terms analytic in the strange quark mass do not uniquely arise
from such loops and
must be fixed by other observables.

In this work we include all terms of the form $m_s \log m_s$
to the axial matrix elements appearing in hyperon decay and
$\beta$-decay used to determine the axial couplings
$F,D,C$ and $H$.
We use these fits to predict matrix
elements relevant for
determining $\Sigma$ , $\Delta s$
and for the interaction of neutrinos with hyperons,
a situation that may be important
at high matter densities \cite{Prak,RPa} .
Unfortunately, higher order SU(3) breaking
contributions can only be
estimated to be of   order $M_K^2/\Lambda_\chi^2\sim 0.25$
(which is not to be confused with
a $25\%$ correction to each matrix element).
Part of the terms at this order (in fact a summation to all orders)
arise from graphs involving the  decuplet of baryon resonances
as intermediate states.
Such contributions  are also present in the
flavour-diagonal axial matrix
elements with the same uncertainty arising from
omission of incalculable terms ${\cal O}(m_s)$ and
higher.
It is clear that our work provides merely an estimate
for the size of SU(3) breaking in these matrix elements,
however, the terms considered here are formally
dominant in the chiral limit.

It is conventional to define the axial matrix
elements of the quarks
in the proton, $|P\rangle$,  via
\begin{eqnarray}
2 s_\mu\ \Delta q & = &
\langle P |\ \overline{q}\gamma_\mu \gamma_5 q\ |P\rangle
\ \ \ ,
\end{eqnarray}
where $q= u, d, s$ denotes the quark flavour, and $s_\mu$ is the
nucleon spin vector.
Any linear combination of the three light quark  neutral axial
currents can be written
in terms of the two diagonal octet generators and the singlet.
In deep-inelastic scattering one measures the matrix
element of the current
\begin{eqnarray}
j_5^\mu & = & \overline{q}\ {\cal Q}^2\ \gamma^\mu\gamma_5 \ q
\ \ \ ,
\end{eqnarray}
in the proton,
where ${\cal Q}$ is the light quark charge matrix, given by
\begin{eqnarray}
{\cal Q} & = & {1\over 3} \left( \matrix{2&0&0 \cr
0&-1&0 \cr 0&0&-1}\right)
\ \ \ .
\end{eqnarray}
In conjunction with a measurement of the matrix elements
of the flavour diagonal currents
\begin{eqnarray}
j_5^{\mu,3} & = &\ \overline{q}\ {\cal O}_3 \gamma_\mu\gamma_5 \ q
\ \ \ \ ,\ \ \ \
j_5^{\mu,8}  =
\ \overline{q}\ {\cal O}_8 \gamma_\mu\gamma_5 \ q
\ \ \ \  ,
\end{eqnarray}
in the proton,
where we use
\begin{eqnarray}
{\cal O}_3 & = & \left(\matrix{1&0&0\cr 0&-1&0\cr 0&0&0}\right)
\ \ \ \ , \ \ \ \
{\cal O}_8  =  \left(\matrix{1&0&0\cr 0&1&0\cr 0&0&-2}\right)
\ \ \ \  ,
\end{eqnarray}
the flavour singlet or alternately the strange quark
contribution may be extracted
via
\begin{eqnarray}
\left(\matrix{1&0&0\cr 0&1&0\cr 0&0&1}\right) & = &
{9\over 2} {\cal Q}^2
- {3\over 4}{\cal O}_3 - {1\over 4}{\cal O}_8
\\
\left(\matrix{0&0&0\cr 0&0&0\cr 0&0&1}\right) & = &
{3\over 2} {\cal Q}^2 - {1\over 4} {\cal O}_3 -
{5\over 12} {\cal O}_8
\ \ \ \ .
\end{eqnarray}

The matrix element of $j_5^{\mu,3}$ in the nucleon
is well determined from nuclear $\beta$-decay via
isospin symmetry, leading to
\begin{eqnarray}
\Delta u\ -\ \Delta d & = & g_A \ =\ 1.2664\pm 0.0065
\ \ \ ,
\end{eqnarray}
where we have neglected isospin breaking effects.
Unfortunately, we cannot use isospin to relate the
matrix element of
$j_5^{\mu,8}$ in the proton
to any other set of physical observables
We must resort to using
flavour SU(3) symmetry as a starting point and
systematically
determine
corrections arising from SU(3) breaking.

Let us begin by discussing the matrix element of the axial
currents in the limit of exact SU(3).
The
matrix elements between baryons in the lowest lying octet
of the axial currents transforming as octets under SU(3)
are described by the following effective lagrange density
\begin{eqnarray}
j_{5,{\rm eff.}}^{\mu,a} & = &
D\ Tr\left[ \overline{B} \ 2s_\mu \{ {\cal O}_a , B\}\right]
\ +\
F\ Tr\left[ \overline{B} \ 2s_\mu [ {\cal O}_a , B ]\right]
\ \ \ ,
\end{eqnarray}
where $B$ is the octet of baryon fields
\begin{eqnarray}
B = \left(
\matrix{\Lambda/\sqrt{6} + \Sigma^0/\sqrt{2} & \Sigma^+ & p\cr
\Sigma^- & \Lambda/\sqrt{6} - \Sigma^0/\sqrt{2} & n\cr
\Xi^- & \Xi^0 & -2 \Lambda/\sqrt{6} }
\right)
\ \ \ \ .
\end{eqnarray}
Also, the matrix element of the singlet current is reproduced by
the lagrange density
\begin{eqnarray}
j_{5,{\rm eff.}}^{\mu,1} & = &
S Tr \left[ \overline{B}\ 2s_\mu\  B \right]
\ \ \ .
\end{eqnarray}

At tree-level we can determine the parameters $F$ and $D$ by
fitting the theoretical
expression, linear in $F$ and $D$, to the observed
rates for
$n\rightarrow p e^-\overline{\nu}_e$,
$\Sigma^-\rightarrow n e^-\overline{\nu}_e$,
$\Xi^-\rightarrow \Lambda e^-\overline{\nu}_e$,
$\Sigma^-\rightarrow\Lambda e^-\overline{\nu}_e$,
$\Xi^-\rightarrow\Sigma^0 e^-\overline{\nu}_e$
and
$\Lambda\rightarrow p e^-\overline{\nu}_e$.
However, one must keep in mind that we expect deviations
between the ``best fit'' and
the experimental results to be at the $\sim 25\%$ level
due to the fact that the
theoretical expressions have been truncated, and terms of
order ${\cal O}\left(
m_s,\ m_s\log m_s, ...\right)$ have been neglected
\cite{JMgg}~.
This includes the fit to the experimentally well measured
value of $g_A$,
equal to $D+F$ in the SU(3) limit (i.e. we naively expect to see
$D+F$ deviate from $g_A$  at the $25\%$ level in the best fit).
In the matrix elements we use to fit the axial couplings
the experimental
uncertainties are much less
than the corresponding theoretical uncertainty.
To determine $F$ and $D$ we minimize a $\chi^2$ function
\begin{eqnarray}
\chi^2 & = & \sum_{\rm data}\
{ ({\rm expt}_i - {\rm theory}_i)^2\over \sigma_{\rm theory}^2}
\ \ \  ,
\end{eqnarray}
where
${\rm expt}_i$ denotes an experimental measurement of an
axial matrix element,
${\rm theory}_i$ denotes its theoretical value for given
values of $F$ and $D$, and
$\sigma_{\rm theory}$ denotes the theoretical uncertainty
which we somewhat
arbitrarily choose to be $\sim 0.2$,
and  equal
for all data points, i.e. an unweighted fit.
This is in contrast to the fit made by Jaffe and Manohar in
\cite{JMgg}
and is a more extreme version of a fit made in \cite{DDJM}\ .
The uncertainties we quote for the couplings $F$ and $D$
are found by requiring that $\chi^2 < \chi^2_{\rm min.} + 2.3$,
corresponding to a $68\%$ confidence interval.
It is clear that this analysis can only provide an estimate
of the uncertainties
as the pattern of breaking will not be uncorrelated for these
processes.
We find that
\begin{eqnarray}
D & = & 0.79\pm 0.10
\cr
F & = & 0.47\pm 0.07
\ \ \ .
\end{eqnarray}
The errors  on $D$ and $F$ are highly correlated and one finds that
the ``best fit'' value for $D+F$ is $1.26\pm 0.08$.
Further, the best for $3F-D$ (the tree-level expression
for the matrix element of
the ${\cal O}_8$ current) is $0.65\pm 0.21$, in agreement
with the central  value of $0.60$ found in \cite{JMgg}.
The values of $F$ and $D$ are in   agreement with those found
in \cite{DDJM}\ except the uncertainties found from our somewhat
{\it ad hoc} procedure are  larger, but they do represent a
reasonable estimate of  the true uncertainties.

A third input required to fix the
individual quark axial matrix elements in the proton is measured in
deep-inelastic scattering
\begin{eqnarray}
2 s_\mu \int_0^1\ dx\ g_1 (x,Q^2) & = &\ {1\over 2}
\left( 1-{\alpha_s (Q^2) \over\pi} \right)
\langle P | \overline{q}\ {\cal Q}^2
\ \gamma_\mu \gamma_5 \ q |P\rangle
\cr
& = & \ {1\over 9} \left( 1-{\alpha_s (Q^2)\over\pi} \right)
\langle P | {3\over 4}\overline{q}\ {\cal O}_3 \
\gamma_\mu \gamma_5 \ q
\ +\  {1\over 4}\overline{q}\ {\cal O}_8 \
\gamma_\mu \gamma_5 \ q
\ +\  \overline{q}\ I \ \gamma_\mu \gamma_5 \ q
|P\rangle
\ \ \ \ ,
\end{eqnarray}
where $I$ is the identity matrix.
The two recent measurements of this quantity are
\begin{eqnarray}
\int_0^1\ dx\ g_1 (x,Q^2 = 3 {\rm GeV}^2)
& = & 0.127\pm 0.004\pm 0.010
\ \ \ ,
\end{eqnarray}
by the E-143 collaboration
\cite{SLACg}
and
\begin{eqnarray}
\int_0^1\ dx\ g_1 (x,Q^2 = 10 {\rm GeV}^2)
& = & 0.136\pm 0.011\pm 0.011
\ \ \ ,
\end{eqnarray}
by the SMC collaboration
\cite{SMCg}.
We choose to use the E-143 measurement at
$Q^2=3 {\rm GeV}^2$ for our evaluations and find
at tree-level
\begin{eqnarray}
\Delta u \ +\ \Delta d\ +\ \Delta s & = & 0.10\pm 0.10
\ =\ \Sigma
\ \ \ \ ,
\end{eqnarray}
which along with the octet matrix elements allows us to
separate the quark contributions
\begin{eqnarray}
\Delta u & = & 0.77\pm 0.04\ \ ,\ \ \Delta d \ = \ -0.49\pm 0.04
\ \ ,\ \ \Delta s \ =\ -0.18\pm 0.09
\ \ \ \ .
\end{eqnarray}
These values are  consistent with the analysis of
Jaffe and Manohar in
\cite{JMgg}.
The $Q^2$ dependence of $\Sigma$ is very weak
\cite{Kod,Ad} (see also \cite{JMgg} and \cite{AVM})
and so we   set $ \Sigma\sim S$.

We can estimate the leading SU(3) breaking to each
axial matrix element in chiral
perturbation theory.
It is of the form $m_s\log m_s$
arising from
the infrared region of hadronic loops involving
$K$'s, $\eta$'s and $\pi$'s and can be computed exactly.
Such loop graphs are divergent and require the
presence of a local counterterm
analytic in the light quark masses which must
be fit to data.
Some  effects of $K$ and $\eta$ loops on strange quark
observables in the nucleon have been
considered previously, e.g. \cite{KHP,MB}~.

For some hyperon decays the axial matrix element is determined from
an experimental measurement of the ratio of vector to axial vector
matrix elements.
The Ademollo-Gatto theorem
\cite{AG}
protects the vector matrix elements from corrections of
the form $m_s \log m_s$, with leading corrections starting
at ${\cal O}(m_s)$
\cite{AL} .
Consequently, at the order to which we are working
we can consistently ignore deviations of the
vector matrix elements
due to SU(3) breaking
and extract the axial matrix elements from the
ratio of axial to vector current matrix elements.

Heavy baryon chiral perturbation theory
\cite{JMa,JMb}
(see also \cite{JMhung})
is used to compute the ${\cal O}(m_s\log m_s)$
corrections to the axial matrix elements.
This technique is sufficiently well known that
we will not go into details in this
work and merely give results of the computation.
The lagrange density for the interaction between
the lowest lying octet and decuplet baryons
of four-velocity $v_\alpha$
with the
pseudo-Goldstone bosons is
\begin{eqnarray}
{\cal L} & = &
Tr[ \overline{B}\ iv\cdot {\cal D}\ B\ ]
\ +\
D Tr[\ \overline{B}\ 2 s_\mu \ \{ A^\mu , B\} ]
\ +\ F Tr[\ \overline{B}\ 2 s_\mu \ [ A^\mu , B ] ]
\\
& - &
\overline{T} \ iv\cdot {\cal D}\ T
\ +  \ \Delta_0 \overline{T}T
\ +\ C \left( \overline{T}^\mu A_\mu B\ +\ h.c.\ \right)
\ +\ H \overline{T}^\mu 2s_\nu A^\nu T_\mu
\ \ \ \  ,
\end{eqnarray}
where ${\cal D}$ is the chiral covariant derivative and
\begin{eqnarray}
A_\mu & = & {i\over 2} \left( \xi\partial_\mu\xi^\dagger
- \xi^\dagger \partial_\mu\xi \right)
\ \ \ ,
\end{eqnarray}
is the axial meson field with
\begin{eqnarray}
\xi & = & exp\left( {i\over f} M \right)
\\
M & = & \left(
\matrix{\eta/\sqrt{6} + \pi^0/\sqrt{2} & \pi^+  & \ K^+\cr
\pi^- & \eta/\sqrt{6} - \pi^0/\sqrt{2} & K^0 \cr
K^- & \overline{K}^0 & -2/\sqrt{6}\eta } \right)
\ \ \ ,
\end{eqnarray}
and $f$ is the meson decay constant.
The axial constants $F,D,C$ and $H$ have been discussed
extensively in the literature
and are seen to be consistent with spin-flavour
SU(6) relations
\cite{JMa,JMb,JMhung,BSSa}.
The mass difference between the decuplet and the
octet baryons is $\Delta_0$.

The matrix element of an axial current with flavour index
$a$ between two octet
baryons states $B_i$ and $B_j$
is given by
\footnote{We have assumed the matrix element
is independent of the invariant mass of the lepton pair.
This is a reasonable approximation as the energy release in these
decays is small.}
\begin{eqnarray}
\langle B_i | j_5^{\mu,a} | B_j \rangle & = &
\overline{U}_i\ \gamma_\mu\gamma_5 U_j
\ \left[
\alpha_{ij}^a\ +\
\left( \beta_{ij}^a - \lambda_{ij} \alpha_{ij}^a \right)
{M_K^2\over 16\pi^2 f^2}\log{\left( M_K^2/\Lambda_\chi^2
\right)}
\ +\ C_{ij}^a (\Lambda_\chi)\ +\ ....
\right]
\ \ \ ,
\end{eqnarray}
where we will take $f$ to be the kaon decay constant
(motivated by previous
experience with such corrections, e.g. \cite{JLMS}),
$f_K = 1.22 f_\pi$,
and $f_\pi = 132\  {\rm MeV}$.
In writing the matrix elements this way we have used the
Gell-Mann-Okubo mass formula
$M_\eta^2 = {4\over 3} M_K^2$ and set $M_\pi = 0$.
The coefficients $\alpha_{ij}^a , \beta_{ij}^a $
for flavour-off-diagonal currents and for $a=8$ in the
proton, along with the wavefunction renormalization
coefficients
$\lambda_{ij}$ have been
computed by Jenkins and Manohar
\cite{JMa,JMb,JMhung}.
The unknown counterterms that contribute at order
${\cal O}(m_s)$ are denoted by
$C_{ij}^a(\Lambda_\chi)$ where we have chosen to
renormalize at the scale
$\mu=\Lambda_\chi$.
As they are unknown quantities, we will set them equal to zero
for our discussions, $C_{ij}^a = 0$.
The coefficients $\alpha_{ij}^a, \beta_{ij}^a$ and
$\lambda_{ij}$
are given in tables I-IV for $\Delta_0=0$.
It is simple to include a non-zero value for the
decuplet-octet mass
difference, $\Delta_0$.
For the vertex graphs involving two decuplet states
and the wavefunction
graphs one makes the replacement
\begin{eqnarray}
M_K^2 \log \left({M_K^2\over \Lambda_\chi^2}\right)
& \rightarrow &
{\cal F}\left( {M_K^2\over \Delta_0} \right)
\\
{\cal F}\left( {M_K^2\over \Delta_0} \right)
& = &
\left( M_K^2 - 2 \Delta_0^2 \right)
\log \left({M_K^2\over \Lambda_\chi^2}
\right)
+ 2 \Delta_0 \sqrt{\Delta_0^2 - M_K^2}
\log \left( { \Delta_0
- \sqrt{\Delta_0^2 - M_K^2 + i \epsilon}
\over \Delta_0
+ \sqrt{\Delta_0^2 - M_K^2 + i \epsilon} }\right)
\ \ \ ,
\end{eqnarray}
and for vertex graphs involving one decuplet
state and one octet state
one makes the replacement
\begin{eqnarray}
M_K^2 \log \left({M_K^2\over \Lambda_\chi^2}\right)
& \rightarrow &
\int_0^1\ dx\
{\cal F}\left( {M_K^2\over  (x \Delta_0 )} \right)
\ \ \ \ .
\end{eqnarray}
Similar replacements occur for the $\eta$ loop graphs.
It was shown by Jenkins and Manohar
\cite{JMa,JMb,JMhung}\ that
it is important to include the decuplet as a dynamical
field otherwise the
natural size of
local counterterms is set by the decuplet-octet
mass splitting and not
$\Lambda_\chi$.
The difference between $\Delta_0\ne 0$ and $\Delta_0=0$
is formally higher order
in the expansion than we are working, however,
setting $\Delta_0\ne 0$ does
allow one to estimate the size of higher order effects.
For our purpose we treat $\Delta_0$ to be the same for all
the decuplet-octet mass splittings and we present results for
$\Delta_0=0\ , 200\ {\rm MeV}$ and $\infty$.
The $\Delta_0=\infty$ theory does not correspond to taking the
$\Delta_0\rightarrow\infty$ limit of
${\cal F}\left( {M_K^2\over \Delta_0} \right)$.
In this limit the function becomes analytic in the light
quark masses
and can be absorbed into a renormalization of higher
order counterterms.
Therefore, the $\Delta_0=\infty$ theory is equivalent to
one without  contributions from the decuplet
(this is also the reason why we can consistently
treat the contribution from $\pi$ loops as negligible).
Also, results for $\Delta_0=300\ {\rm MeV}$ are little
different from those for
$\Delta_0=200\ {\rm MeV}$.

\begin{table}
\begin{tabular}{ccc}
& \multicolumn{2}{c}  { {\em coefficients} } \\
{\em process} & {$\alpha_{ij}^3$} &  {$\beta_{ij}^3$}
\\   \tableline
\rule{0cm}{0.5cm}
$p\rightarrow p$
&$D+F$
&${4\over 9}\left( D^3 + D^2F + 3DF^2 - 9 F^3\right) - D - F
-{20\over 81} C^2 H + {4\over 9} C^2 (F+3 D)$
\\  \tableline
\rule{0cm}{0.5cm}
$\Sigma^+\rightarrow \Sigma^+$
&$2 F$
&$ -2F - {2\over 9} F (9F^2-D^2) - {50\over 27} C^2 H
+{8\over 3} C^2 ({13\over 9} D - {1\over 3} F ) $
\\  \tableline
\rule{0cm}{0.5cm}
$\Xi^0\rightarrow \Xi^0$
&$F-D$
&$D-F - {4\over 9}\left( D^3 - D^2F + 3 D F^2 + 9 F^3 \right)
- {40\over 81} C^2 H - {8\over 3} C^2 ( {7\over 18} D
+ {3\over 2} F ) $
\\  \tableline
\rule{0cm}{0.5cm}
$\Lambda\rightarrow \Sigma^0$
&${2\over\sqrt{3}}D$
&$-{1\over\sqrt{3}} \left[ 2 D
+ {2\over 9}D (9F^2-17D^2) + {10\over 27} C^2H
- {16\over 3} C^2 (D+F)
\right] $
\\
\end{tabular}
\vskip 0.5cm
\caption{The coefficients $\alpha_{ij}^3$ and
$\beta_{ij}^3$ for the
flavour-diagonal axial matrix elements ($\Delta_0=0$).
The remaining matrix elements are related by isospin to
those in the table.}
\end{table}

\begin{table}
\begin{tabular}{ccc}
& \multicolumn{2}{c}  { {\em coefficients} } \\
{\em process} & {$\alpha_{ij}^8$} &  {$\beta_{ij}^8$}
\\   \tableline
\rule{0cm}{0.5cm}
$p\rightarrow p$
&$3F-D$
&$3D-9F - {2\over 9}\left( 11D^3 -27 D^2F -27DF^2 +27 F^3\right)
+ 4 C^2 (D-F)$
\\  \tableline
\rule{0cm}{0.5cm}
$\Lambda\rightarrow \Lambda$
&$-2 D$
&$6D - {2\over 9}D( 27F^2-11D^2 )
+ {10\over 9}C^2H  + {8\over 3} C^2 (D-3F)$
\\  \tableline
\rule{0cm}{0.5cm}
$\Sigma^+\rightarrow \Sigma^+$
&$2 D $
&$ -6D + {2\over 9} D (D^2+63F^2) - {10\over 9} C^2 H
+ {8\over 3} C^2 ({7\over 3} D + F )   $
\\  \tableline
\rule{0cm}{0.5cm}
$\Xi^0\rightarrow \Xi^0$
&$-D-3F$
&$3D + 9F - {2\over 9}\left(11 D^3 +27 D^2F
-27 D F^2 -27 F^3 \right)
 - {8\over 3} C^2 ( {13\over 6} D + {7\over 2} F
 - {10\over 9}H )  $
\\
\end{tabular}
\vskip 0.5cm
\caption{The coefficients $\alpha_{ij}^8$ and
$\beta_{ij}^8$ for the
flavour-diagonal axial matrix elements ($\Delta_0=0$).
The remaining matrix elements are related by
isospin to those in the table.}
\end{table}

\begin{table}
\begin{tabular}{ccc}
& \multicolumn{2}{c}  { {\em coefficients} } \\
{\em process} & {$\alpha_{ij}^1$} &  {$\beta_{ij}^1$}
\\   \tableline
\rule{0cm}{0.5cm}
$p\rightarrow p$
&$S$
&$-  S ( 5 F^2 + {17\over 9}D^2 - {10\over 3}FD)
- {\cal T} {5\over 9} C^2
$
\\  \tableline
\rule{0cm}{0.5cm}
$\Lambda\rightarrow \Lambda$
&$S$
&$-  S ( 6 F^2 + {14\over 9}D^2  )
- {\cal T} {10\over 9} C^2 $
\\  \tableline
\rule{0cm}{0.5cm}
$\Sigma\rightarrow \Sigma$
&$S$
&$ -  S ( 2 F^2 + {26\over 9}D^2  )
- {\cal T} {70\over 27} C^2    $
\\  \tableline
\rule{0cm}{0.5cm}
$\Xi\rightarrow \Xi$
&$S$
&$-  S ( 5 F^2 + {17\over 9}D^2 + {10\over 3}FD)
- {\cal T} {65\over 27} C^2   $
\\
\end{tabular}
\vskip 0.5cm
\caption{The coefficients $\alpha_{ij}^1$ and
$\beta_{ij}^1$ for the
flavour singlet axial matrix elements ($\Delta_0=0$).
}
\end{table}

\begin{table}
\begin{tabular}{cc}
{\em process} & {$\lambda_{ij}$}
\\   \tableline
\rule{0cm}{0.5cm}
$N\rightarrow N$
&${17\over 3} D^2 + 15 F^2 - 10 DF + C^2$
\\  \tableline
\rule{0cm}{0.5cm}
$\Sigma\rightarrow \Sigma$
&${26\over 3} D^2  + 6 F^2 + {14\over 3} C^2$
\\  \tableline
\rule{0cm}{0.5cm}
$\Lambda\rightarrow \Lambda$
&${14\over 3} D^2  + 18 F^2 + 2 C^2 $
\\  \tableline
\rule{0cm}{0.5cm}
$\Xi\rightarrow \Xi$
&${17\over 3} D^2 + 15 F^2 + 10 DF + {13\over 3}C^2$
\\  \tableline
\rule{0cm}{0.5cm}
$\Lambda\rightarrow \Sigma$
&${20\over 3} D^2 + 12 F^2   + {10\over 3} C^2$
\\
\end{tabular}
\vskip 0.5cm
\caption{The wavefunction renormalization
coefficients $\lambda_{ij}$ ($\Delta_0=0$). }
\end{table}

Notice that at this order we are forced to introduce an
unknown parameter
${\cal T}$, the matrix element of the singlet axial
current in the  decuplet, or
equivalently, the strange content of the $\Delta$.
It arises in the loop graphs involving decuplet intermediate states
(there is no octet to decuplet transition induced by the singlet),
\begin{eqnarray}
j_5^{\mu,1} ({\bf 10}) & = &
{\cal T}\  \overline{T}_\alpha^{abc}\ 2 s_\mu\ T_{abc}^\alpha
\ \ \  .
\end{eqnarray}
The value of this constant is unknown and for our
calculations we set ${\cal T}=0$ (setting
${\cal T} = S$ gives virtually identical results).
However, this quantity does provide a problem for a
systematic inclusion of higher
order corrections to the SU(3) limit.
Physically one extracts a linear combination of $S$ and
${\cal T}$ at one-loop order and the same
linear combination enters in all appropriate
observables in the nucleon
sector at this order.
However,
when considering matrix elements between strange hyperons
a different linear
combination of $S$ and ${\cal T}$ will enter.

The axial couplings of the decuplet $C$ and $H$
first contribute to the axial matrix
elements of the octet baryons at loop-level and
hence cannot be well constrained
from the semileptonic decays alone.
In addition to the $\beta$-decay and the hyperon
decay used for the
tree-level fit, we require that the couplings
reproduce the strong decays of
the  $\Delta, \Sigma^*$ and the $\Xi^*$.
Expressions for these rates at
${\cal O}(m_s\log m_s)$ can be found in
\cite{BSSa} (tree-level extractions
would be sufficient at this order ) .
The  procedure for the tree-level fitting was
applied to the loop-level fitting, except that we fit four
coupling constants instead of the two at tree-level (i.e.
$\chi^2 < \chi_{\rm min}^2 + 4.7$).
Best fit values for the axial coupling constants are
shown in table V and
they are consistent with previous extractions.
The fits to the semileptonic decay matrix elements
both at tree-level and one-loop
level are shown in table VI.
Differences between the tree-level and loop-level
fits to the semileptonic matrix elements are not great.
Neutral current axial matrix elements are
estimated at leading order in
SU(3) breaking and we present the estimates for
${\cal O}_3$,
${\cal O}_8$ and the singlet current
for each of the octet baryons in tables VII-IX.

\begin{table}
\begin{tabular}{cccc}
& \multicolumn{3}{c}  { {\em Axial Coupling Constants} }  \\
{\em coupling} & {$\Delta_0=0$} &  {$\Delta_0=200 {\rm MeV}$}
& {$\Delta_0=\infty$}
\\   \tableline
\rule{0cm}{0.5cm}
$D$
&$0.64\pm 0.05$
&$0.64 \pm 0.06$
&$0.59\pm0.06$
\\  \tableline
\rule{0cm}{0.5cm}
$F$
&$0.42\pm 0.04$
&$0.34 \pm 0.04$
&$0.34\pm 0.04$
\\  \tableline
\rule{0cm}{0.5cm}
$|C|$
&$1.39\pm 0.06$
&$1.37\pm 0.05$
&$1.37\pm 0.06$
\\  \tableline
\rule{0cm}{0.5cm}
$H$
&$-2.7\pm 0.6$
&$-2.7 \pm 0.5$
&$-2.8\pm 0.5$
\\
\end{tabular}
\vskip 0.5cm
\caption{Loop-level axial coupling constants for
$\Delta_0=0, \Delta_0 = 200 {\rm
MeV}$ and $\Delta_0=\infty$.}
\end{table}

\begin{table}
\begin{tabular}{ccccc}
& \multicolumn{3}{c}  { {\em Axial Matrix Elements} }  \\
{\em process} & {\em tree-level} &  {\em loop-level\ $^a$}
&  {\em loop-level\ $^b$}& {\em experimental}\cite{DDJM,PDG}
\\   \tableline
\rule{0cm}{0.5cm}
$n\rightarrow p$
&$1.26\pm 0.08$
&$1.24\pm 0.11$
&$1.16 \pm 0.09$
&$1.2664\pm 0.0065$
\\  \tableline
\rule{0cm}{0.5cm}
$\Sigma^-\rightarrow n$
&$0.31\pm 0.10$
&$0.35\pm 0.13$
&$0.31\pm 0.1$
&$0.341\pm 0.015$
\\  \tableline
\rule{0cm}{0.5cm}
$\Xi^-\rightarrow \Lambda$
&$0.27\pm 0.09$
&$0.25\pm 0.14$
&$0.29\pm 0.10$
&$0.306\pm 0.061$
\\  \tableline
\rule{0cm}{0.5cm}
$\Lambda\rightarrow p$
&$-0.90\pm 0.07$
&$-1.01\pm 0.11$
&$-0.98\pm 0.09$
&$-0.890\pm 0.015$
\\  \tableline
\rule{0cm}{0.5cm}
$\Sigma^-\rightarrow \Lambda$
&$0.64\pm 0.06$
&$0.64\pm 0.06$
&$0.57\pm 0.06$
&$0.602\pm 0.014$
\\  \tableline
\rule{0cm}{0.5cm}
$\Xi^-\rightarrow \Sigma^0$
&$0.89\pm 0.06$
&$0.89\pm 0.06$
&$1.00\pm 0.09$
&$0.929\pm 0.112$
\\  \tableline
\rule{0cm}{0.5cm}
$\Delta\rightarrow N$
&$-1.70\pm 0.07$
&$-1.76\pm 0.13$
&$-1.75\pm 0.11$
&$-2.04\pm 0.01$
\\  \tableline
\rule{0cm}{0.5cm}
$\Sigma^*\rightarrow \Lambda$
&$-1.70\pm 0.07$
&$-1.76\pm 0.14$
&$-1.77\pm 0.12$
&$-1.71\pm 0.03$
\\  \tableline
\rule{0cm}{0.5cm}
$\Sigma^*\rightarrow \Sigma$
&$-1.70\pm 0.07$
&$-1.50\pm 0.18$
&$-1.52\pm 0.15$
&$-1.60\pm 0.13$
\\  \tableline
\rule{0cm}{0.5cm}
$\Xi^*\rightarrow \Xi$
&$-1.70\pm 0.07$
&$-1.64\pm 0.12$
&$-1.65\pm 0.09$
&$-1.42\pm 0.04$
\\
\end{tabular}
\vskip 0.5cm
\caption{Tree- and loop-level evaluations of
matrix elements of the axial current.
Superscript \ $^{a,b}$\  denote $\Delta_0= 0$ and $200\ {\rm MeV}$
respectively.}
\end{table}

\begin{table}
\begin{tabular}{ccccc}
& \multicolumn{4}{c}  { ${\cal O}_3$ } \\
{\em process} & {\em tree-level}
&  {\em loop-level\ $^a$}
& {\em loop-level\ $^b$} &  {\em loop-level\ $^c$}
\\   \tableline
\rule{0cm}{0.5cm}
$\Sigma^+\rightarrow \Sigma^+$
&$0.95\pm 0.12$
&$0.70\pm 0.23$
&$1.09\pm 0.12$
&$0.98\pm 0.14$
\\  \tableline
\rule{0cm}{0.5cm}
$\Xi^0\rightarrow\Xi^0$
&$-0.31\pm 0.10$
&$-0.35\pm 0.17$
&$-0.18\pm 0.10$
&$-0.36 \pm 0.13$
\\
\end{tabular}
\vskip 0.5cm
\caption{Tree-level and loop-level estimates of  the
matrix elements of the $ {\cal O}_3 $
axial current.
The matrix element in the proton
is not shown as it is fixed by isospin  to $ g_A $ .
Similarly, the matrix element for the $\Lambda-\Sigma$
transition is not shown as it is related to
the matrix element for $\Sigma^-\rightarrow \Lambda$
by isospin.
Also, the matrix element between
$ \Lambda $ states  vanishes by isospin.
Superscripts  $ { }^{a, b, c} $
denote
$ \Delta_0 = 0$,  $200 \  {\rm   MeV}  $  and $\infty $
respectively.
}
\end{table}

\begin{table}
\begin{tabular}{ccccc}
& \multicolumn{4}{c}  { ${\cal O}_8$ } \\
{\em process} & {\em tree-level}
&  {\em loop-level\ $^a$}
& {\em loop-level\ $^b$} &  {\em loop-level\ $^c$}
\\   \tableline
\rule{0cm}{0.5cm}
$p\rightarrow p$
&$0.65\pm 0.21$
&$0.78\pm 0.24$
&$0.45\pm 0.20 $
&$0.60\pm 0.22 $
\\  \tableline
\rule{0cm}{0.5cm}
$\Sigma^+\rightarrow \Sigma^+$
&$1.56\pm 0.15$
&$1.63\pm 0.26$
&$1.58\pm 0.18 $
&$1.78\pm 0.23 $
\\  \tableline
\rule{0cm}{0.5cm}
$\Lambda\rightarrow\Lambda$
&$-1.56\pm 0.15$
&$-1.83\pm 0.28$
&$-2.08\pm 0.20$
&$-1.88\pm 0.21 $
\\  \tableline
\rule{0cm}{0.5cm}
$\Xi^0\rightarrow\Xi^0$
&$-2.21\pm 0.17$
&$-2.31 \pm 0.40$
&$-2.79\pm 0.30 $
&$-2.81\pm 0.35 $
\\
\end{tabular}
\vskip 0.5cm
\caption{  Tree-level and loop-level estimates of the
matrix elements of the $ {\cal O}_8 $ axial current.
The matrix element between $\Lambda$ and $\Sigma$ states
vanishes by isospin.
Superscripts $^{a,b,c}$ denote
$ \Delta_0 = 0, 200 \   {\rm MeV}  ,  \infty  $ respectively.
}
\end{table}

\begin{table}
\begin{tabular}{ccccc}
& \multicolumn{4}{c}  { $I$  } \\
{\em process} & {\em tree-level}
&  {\em loop-level\ $^a$}
& {\em loop-level\ $^b$} &  {\em loop-level\ $^c$}
\\   \tableline
\rule{0cm}{0.5cm}
$p\rightarrow p$
&$0.10\pm 0.11$
&$0.08\pm 0.12$
&$0.16\pm 0.11$
&$0.11\pm 0.13$
\\  \tableline
\rule{0cm}{0.5cm}
$\Sigma^+\rightarrow\Sigma^+$
&$0.10\pm 0.11$
&$0.13\pm 0.19$
&$0.15\pm 0.10$
&$0.14\pm 0.15$
\\  \tableline
\rule{0cm}{0.5cm}
$\Lambda\rightarrow\Lambda$
&$0.10\pm 0.11$
&$0.10\pm 0.16$
&$0.18\pm 0.13$
&$0.13\pm 0.15$
\\  \tableline
\rule{0cm}{0.5cm}
$\Xi^0\rightarrow\Xi^0$
&$0.10\pm 0.11$
&$0.14\pm 0.22$
&$0.18\pm 0.13$
&$0.16\pm 0.18$
\\
\end{tabular}
\vskip 0.5cm
\caption{Tree-level and loop-level estimates of the
matrix elements of the singlet axial current extracted from the
E143 measurement of $ \int dx\ g_1(x)\ =\ 0.127\pm 0.004\pm0.010$.
The matrix element between $\Lambda$ and $\Sigma$ states
vanishes by isospin.
Superscripts $^{a,b,c}$ denote
$\Delta_0 = 0,200\ {\rm MeV},\infty$ respectively.
We have set ${\cal T} = 0$ in the loop-level calculations.}
\end{table}

The loop-level extractions of the quark contributions to
the proton spin are
shown in table X, along with the tree-level result.
It is evident that the up and down quark
contributions are insensitive
to the SU(3) breaking.
In contrast, the strange quark content is very sensitive to
the breaking, however, all
the determinations agree within the
uncertainties.
Further, the matrix element of the singlet
current in the proton
extracted from the E-143 measurement of
$\int dx\ g_1(x)\ =\ 0.127\pm 0.004 \pm 0.010$
appears to be compatible with zero in each of the
determinations,
as it is at tree-level.
If instead one used the SMC measurement of
$\int dx\ g_1(x)\ =\ 0.136\pm 0.011 \pm 0.011$
the magnitude of $\Sigma$ is increased by
$\sim 50\%$.
Our loop analysis of the matrix element of
${\cal O}_8$ in the proton
is in disagreement with the analysis of
Dai {\it et al}\cite{DDJM}\ .
In the large-$N_c$ limit they find a value of
$0.27\pm 0.09$,
which is smaller by a factor of two
than our estimates although we do
have a large uncertainty.

\begin{table}
\begin{tabular}{ccccc}
& \multicolumn{4}{c}
{\em Matrix elements of the light quark axial currents
 } \\
{\em quark flavour} & {\em tree-level}
&  {\em loop-level\ $^a$}
& {\em loop-level\ $^b$} &  {\em loop-level\ $^c$}
\\   \tableline
\rule{0cm}{0.5cm}
$\Delta u$
&$0.77\pm 0.04 $
&$0.79\pm 0.04$
&$0.76\pm 0.04$
&$0.77\pm 0.04$
\\  \tableline
\rule{0cm}{0.5cm}
$\Delta d$
&$-0.49\pm 0.04$
&$-0.48\pm 0.04$
&$-0.51\pm 0.04$
&$-0.50\pm 0.04$
\\  \tableline
\rule{0cm}{0.5cm}
$\Delta s$
&$-0.18\pm 0.09$
&$-0.23\pm 0.10$
&$-0.10\pm 0.09$
&$-0.16\pm 0.10$
\\
\end{tabular}
\vskip 0.5cm
\caption{Tree-level and loop-level estimates of the
individual quark contributions
to the proton spin extracted from the
E-143 measurement of
$\int dx\ g_1(x)\ =\ 0.127\pm 0.004 \pm 0.010$~.
Superscripts $^{a,b,c}$ denote
$\Delta_0 = 0,200\ {\rm MeV},\infty$ respectively.
We have set ${\cal T} = 0$ in the loop-level calculations.}
\end{table}

It is useful to understand what situation must arise
in order to recover the naive quark model estimate of
$\Sigma \sim +0.58$.
We find that if $D = 0.66$, $F=0.37$,
$C=-1.4$ and $H=-2.6$, then one can reproduce
most axial couplings arising in
semileptonic rates reasonable well
except for $\Sigma^-\rightarrow n$,
which would have to be $0.50$ compared with
$0.341\pm 0.015$ observed
and
$\Xi^-\rightarrow\Lambda$,
which would have to be $0.09$ compared with
$0.306\pm 0.061$ observed.
Unless the experimental determinations are
many standard deviations away from the true value of
these axial couplings it appears unlikely
that the naive quark model value of $\Sigma$ will
arise.

We should remind ourselves that the measurements planned
to be made at Jefferson
Laboratory of the parity violating component of $ep$
interactions
and LSND running at Los Alamos measuring $\nu p$ scattering
(see  \cite{MDDPKB} for a comprehensive review)
circumvent the
need to use SU(3) symmetry to extract the
strange content of the nucleon and hence will not rely upon
the estimates made here.
The axial current that couples to the $Z^0$ has
flavour structure
\begin{eqnarray}
j_5^{\mu, Z} & = & \overline{q}\ {\cal O}_Z\
\gamma^\mu\gamma_5 \ q
\\
{\cal O}_Z & = & {\cal O}_3
+ {1\over 3}{\cal O}_8 - {1\over 3} I
\\
& = & {\cal O}_3 -
\left(\matrix{0&0&0\cr 0&0&0\cr 0&0&1}\right)
\ \ \ \ ,
\end{eqnarray}
and as
the matrix element of ${\cal O}_3$ in the proton
is known from $g_A$ by
isospin symmetry a measurement of the $Z^0$
axial coupling will yield
the strange quark content of the nucleon directly.
It would appear from our somewhat primitive analysis
of SU(3) breaking that
the $Z^0$ measurements are the key to determining the
strange quark content of
the nucleon.

As an aside we
consider the analogue of the strange
quark content of the nucleon for the other
baryons in the octet.
Such quantities could be  the ``up-quark'' content
of the $\Xi^-$ (with flavour
quantum numbers $ssd$) or the ``down-quark'' content
of the $\Sigma^+$ (with
flavour quantum numbers $uus$).
In the limit of exact SU(3) one can find the
individual quark contributions by
large SU(3) transformations.
For instance under $s\leftrightarrow u$ we have
$p\leftrightarrow \Xi^-$ and
hence we expect that the up-quark content of the
$\Xi^-$ is equal to the
strange quark content of the nucleon.
Similarly, under $s\leftrightarrow d$ we
have $p\leftrightarrow \Sigma^+$ and we expect that
the down-quark content of
the $\Sigma^+$ is the same as the strange quark
content of the nucleon.
We can investigate the effects of the
${\cal O}(m_s \log m_s)$ SU(3) breaking
terms on these relations simply from our
above analysis (we use the $\Delta_0=0$ results).
We find that for the $\Xi^-$ at loop-level
\begin{eqnarray}
\Delta u_\Xi & = & -0.18\pm 0.14
\ \ ,\ \
\Delta d_\Xi = -0.50\pm 0.10
\ \ ,\ \
\Delta s_\Xi = 0.83\pm 0.12
\ \ \ ,
\end{eqnarray}
and for the $\Sigma^+$ at loop-level
\begin{eqnarray}
\Delta u_\Sigma & = & 0.68\pm 0.12
\ \ ,\ \
\Delta d_\Sigma = 0.05\pm 0.12
\ \ ,\ \
\Delta s_\Sigma = -0.49\pm 0.09
\ \ \ .
\end{eqnarray}
We see that the ``wrong''-quark content
is about the same for
each baryon and is consistent with the
results seen in the
nucleon sector alone.
We may make a connection with the nonleptonic
interactions between  octet
baryons and the pseudo-Goldstone bosons.
It was realized in \cite{DLSS,KS}
that a non-zero strange axial matrix element
in the nucleon
may impact nuclear parity violation.
Non-strange operators are suppressed
by custodial symmetries of the
standard model of electroweak interactions
in the limit
$\sin^2\theta_w\rightarrow 0$,
while strange operators are not.
The strangeness changing four-quark
interaction (ignoring strong interaction
corrections) is
\begin{eqnarray}
{\cal H} & = & {G_F\over\sqrt{2}} V_{us}V^\dagger_{ud}
\
\overline{u}\gamma^\mu (1-\gamma_5) s\
\overline{d}\gamma_\mu (1-\gamma_5) u
\ \ \ ,
\end{eqnarray}
and naively one might not expect this operator
to contribute to the weak
coupling $\Xi^- \Xi^- K^0$, as there are no up
quarks in any of the hadrons.
However,  SU(3) symmetry relations arising from
the observed octet enhancement
in these nonleptonic decays gives $S$ and
$P$ wave amplitudes
\begin{eqnarray}
{\cal A}^{(S)} & = & {1\over f} (h_D+h_F)
\\
{\cal A}^{(P)} & = & {(D+F)(h_D+h_F)\over f
(M_\Xi-M_\Sigma)}\ 2S\cdot k
\ \ \ ,
\end{eqnarray}
where $k$ is the outgoing meson momentum and
$h_D$ and $h_F$ are two constants,
determined to  be
$h_D = (-0.58\pm 0.21) G_F M_\pi^2 f$ and
$h_F = (+1.40\pm 0.12) G_F M_\pi^2 f$ at tree-level
\cite{EJ}.
One can also compute these amplitudes in the
factorization limit giving
\begin{eqnarray}
{\cal A}^{(S)}_{\rm fact} & = & 0
\\
{\cal A}^{(P)}_{\rm fact} & = & {G_F\over \sqrt{2}}\
 V_{us}V^\dagger_{ud}
\ f \ (\Delta u_\Xi)\  2 S\cdot k
\ \ \  ,
\end{eqnarray}
where $\Delta u$ is the up quark contribution to
the $\Xi^-$ spin.
In order to reproduce the P-wave amplitude computed
via octet enhancement we
require $\Delta u_\Xi \sim 0.05$, a value that is
encompassed by  our determination.
This suggests that the up-quark content of the
$\Xi^-$ could lead to a
counterterm  for the nonleptonic vertex,
$\sim {\cal A}^{(P)}_{\rm fact}$,
that is
the same size if not
larger than the vertex resulting from the
baryon pole graph,
${\cal A}^{(P)}$~.

In systems of density comparable to or greater than that
of nuclear matter
such as arise in ``neutron stars'', the exact composition
of the matter is far from
certain.
The strange quark is guaranteed to play a role at high
enough density, but the
question of at what density it becomes important depends
crucially on the strong
interactions between the nucleons, the
strange hyperons and the mesons.
If indeed it is energetically favoured for strange
baryons to be present in
significant number densities
then it is necessary to know the
interactions of neutrinos with these baryons in
order to construct a reasonable
model for the evolution of some dense matter systems
\cite{Prak,RPa} .
We  present estimates of the axial matrix
elements for $Z^0$ interactions between hyperons
in the lowest lying octet, $C_A$,
in table XI.
It is clear that some matrix elements are more susceptible
to large SU(3) breaking corrections than others,
at least for the corrections that we
could estimate.
In particular matrix elements for the $\Sigma^-$ and $ \Xi^-$
appear to be particularly unreliable,
with large deviations from the tree-level estimates likely.

\begin{table}
\begin{tabular}{ccccc}
& \multicolumn{3}{c}  { $C_A$ } \\
{\em process} &  {$\Delta s=0$ {\em tree-level} }
& {\em tree-level}
&  {\em loop-level \ $^a$} &  {\em loop-level \ $^b$}
\\   \tableline
\rule{0cm}{0.5cm}
$p\rightarrow p$
&$D+F = 1.26$
&$1.43\pm 0.10$
&$1.50\pm 0.10$
&$1.36\pm 0.09$
\\  \tableline
\rule{0cm}{0.5cm}
$n\rightarrow n$
&$-(D+F) = -1.26$
&$-1.09\pm 0.10$
&$-1.04\pm 0.10$
&$-1.17\pm 0.09$
\\  \tableline
\rule{0cm}{0.5cm}
$\Lambda\rightarrow\Lambda$
&$-(F+D/3) = -0.73$
&$-0.56\pm 0.07$
&$-0.64\pm 0.11$
&$-0.75\pm 0.09$
\\  \tableline
\rule{0cm}{0.5cm}
$\Sigma^+\rightarrow\Sigma^+$
&$D+F = 1.26$
&$1.44\pm 0.13$
&$1.20\pm 0.25$
&$1.57\pm 0.15$
\\  \tableline
\rule{0cm}{0.5cm}
$\Sigma^0\rightarrow\Sigma^0$
&$D-F = 0.34$
&$0.46\pm 0.04$
&$0.50\pm 0.09$
&$0.48\pm 0.06$
\\  \tableline
\rule{0cm}{0.5cm}
$\Sigma^-\rightarrow\Sigma^-$
&$D-3F = -0.58$
&$-0.46\pm 0.13$
&$-0.19\pm 0.20$
&$-0.61\pm 0.12$
\\  \tableline
\rule{0cm}{0.5cm}
$\Xi^0\rightarrow\Xi^0$
&$-(D+F) = -1.26$
&$-1.07\pm 0.11$
&$-1.17\pm 0.20$
&$-1.17\pm 0.13$
\\ \tableline
\rule{0cm}{0.5cm}
$\Xi^-\rightarrow\Xi^-$
&$D-3F = -0.58$
&$-0.47\pm 0.13$
&$-0.46\pm 0.23$
&$-0.81\pm 0.16$
\\ \end{tabular}
\vskip 0.5cm
\caption{Tree-level and loop-level evaluations of the
matrix elements of the  neutral  axial current
coupling to the  $Z^0$ .
Superscripts $^{a,b}$ denote
$\Delta_0 = 0, 200\  {\rm MeV}$ respectively.
Isospin relates the matrix element for
$\Sigma^-\rightarrow\Lambda$
to the value of $C_A$ for the $\Lambda-\Sigma^0$
transition,
giving $C_A = 0.85\pm 0.02$.}
\end{table}

In conclusion, we have computed the leading, model independent
SU(3) breaking
contributions to the matrix
elements of axial current with flavour structure
${\cal O}_3$, ${\cal O}_8$ and the
flavour singlet.
We find that there is a large uncertainty in some matrix elements,
and this is probably an indication of comparable uncertainty
in all matrix elements from terms we cannot compute.

It is the matrix element of ${\cal O}_8$ in the proton that
presently impacts the determination of the $\Delta s $ and $\Sigma$
in the proton.
We find that both quantities are sensitive to SU(3) breaking
(in disagreement with  \cite{LL} where the
impact of SU(3) violation on $\Sigma$ was claimed to be small)
and we estimate them to  lie in the intervals
$-0.1\  \lsim\  \Sigma\  \lsim\ +0.3$
 and $-0.35\  \lsim\  \Delta s\  \lsim\ 0$
from the E-143 measurement of
$\int dx\ g_1(x)\ =\ 0.127\pm 0.004 \pm 0.010$~.
The upper limit of this range for $\Sigma$
is still much less than the
naive quark model estimate of $+0.58$
(using the SMC value for
$\int dx\ g_1(x)\ =\ 0.136\pm 0.011 \pm 0.011$
the upper limit of $\Sigma$ becomes $\sim 0.35$)~.
Somewhat more pessimistically, we clearly
demonstrate that there is a
large theoretical impediment to making a more
precise determination of $\Sigma$ and $\Delta s$
from better measurements of $g_1 (x)$.
It appears that improvement can only occur from measurements
of the $Z^0$ coupling to nucleons.

\vskip 1.5cm

\centerline{\bf Acknowledgements}

This work was stimulated by discussion with M. Prakash
and S. Reddy
at the Institute for Nuclear Theory at the University
of Washington in
Seattle.
We would like to thank D. Kaplan,
A. Manohar, R. Springer, M. Prakash, Gerry Miller
and H. Robertson for helpful comments.
This work is supported by the Department of Energy.

\begin{figure}
\epsfxsize=6cm
\hfil\epsfbox{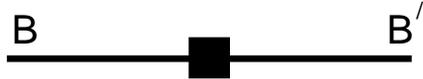}\hfill
\caption{Tree-level contribution to the axial matrix element.
The solid square denotes the insertion of the axial current.
The labels $B$ and $B^\prime$ denote the incoming and outgoing
octet baryons respectively.
The dashed line denotes a pseudo-Goldstone boson.}
\label{treefig}
\end{figure}

\begin{figure}
\epsfxsize=10cm
\hfil\epsfbox{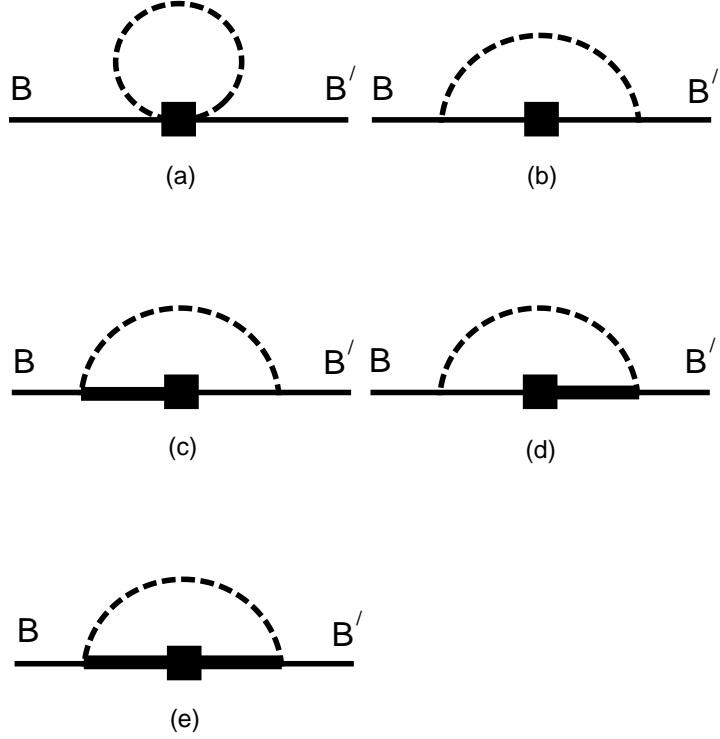}\hfill
\caption{Loop-level contribution to the axial matrix element.
The solid square denotes the insertion of the axial current.
The labels $B$ and $B^\prime$ denote the incoming and outgoing
octet baryons respectively.
The dashed line denotes a pseudo-Goldstone boson.
The thicker lines denote decuplet baryon propagators.
Graphs of the type (a), (c) and (d) do not arise in the
matrix element of the singlet current at one-loop.}
\label{loopfig}
\end{figure}

\begin{figure}
\epsfxsize=10cm
\hfil\epsfbox{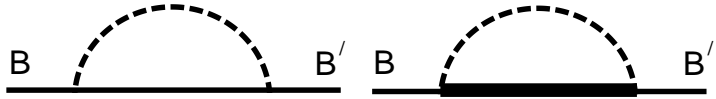}\hfill
\caption{Loop-level wavefunction renormalization
contributions to the axial matrix element.
The solid square denotes the insertion of the axial current.
The labels $B$ and $B^\prime$ denote the incoming and outgoing
octet baryons respectively.
The dashed line denotes a pseudo-Goldstone boson.
The thicker lines denote decuplet baryon propagators.}
\label{wavefig}
\end{figure}

\end{document}